\documentclass[conference]{IEEEtran}
\usepackage{multirow}
\usepackage{threeparttable}
\usepackage{color}
\usepackage{graphicx}
\usepackage{algorithm}
\usepackage{algpseudocode}
\usepackage{subfigure}
\usepackage{textcomp,booktabs}
\usepackage{colortbl}
\usepackage{array}
\usepackage{calc}
\usepackage{hhline,tabu}
\usepackage{amsmath}
\usepackage{setspace}
\usepackage{siunitx}
\usepackage{breqn}
\usepackage{bm}

\usepackage[perpage,symbol]{footmisc}
\setfnsymbol{wiley}
\newcommand{\tabincell}[2]{\begin{tabular}{@{}#1@{}}#2\end{tabular}}

\begin{document}
%
\title{The Outbreak Evaluation of COVID-19 in Wuhan District of China}


\author{\IEEEauthorblockN{Yimin Zhou$^{1,2}$, Zuguo Chen$^{1,3}$, Xiangdong Wu$^1$, Zengwu Tian$^1$, Liang Cheng$^1$, Lingjian Ye$^1$}
\IEEEauthorblockA{$^1$Shenzhen Institutes of Advanced Technology, Chinese Academy of Sciences, \\1068 Xueyuan Avenue, Xili University Town, Shenzhen, China, 518055\\}
\IEEEauthorblockA{$^2$Xi'an University of Posts \& Telecommunications, Xi'an, China, 710121}
\IEEEauthorblockA{$^3$Hunan University of Science and Technology, Hunan, China,  411201}
Email: ym.zhou@siat.ac.cn

}


\maketitle

\begin{abstract}
There were 27 novel coronavirus pneumonia cases found in Wuhan, China in December 2019, named as 2019-nCoV temporarily and COVID-19 formally by WHO on $11^{th}$ February, 2020. In December 2019 and January 2020, COVID-19 has spread in large scale among the population, which brought terrible disaster to the life and property of the Chinese people. In this paper, we will first analyze the feature and pattern of the virus transmission, and discuss the key impact factors and uncontrollable factors of epidemic transmission based on public data. Then the virus transmission can be modelled and  used for the inflexion and extinction period of epidemic development so as to provide theoretical support for the Chinese government in the decision-making of epidemic prevention and recovery of economic production. Further, this paper demonstrates the effectiveness of the prevention methods taken by the  Chinese government such as multi-level administrative region isolation. It is of great importance and practical significance for the world to deal with public health emergencies.
\end{abstract}

\footnote{The data in the paper are collected till 24:00 $17^{th}$ Feb 2020.} 
\footnotemark[1]

%
\IEEEpeerreviewmaketitle

\section{Introduction}
The outbreak of novel coronavirus pneumonia (NCP) in Wuhan at the end of 2019 has been spreading rapidly, which was designated as ``public health emergencies of international concern" by World Health Organization (WHO). In just one month from the beginning of January 2020, COVID-19 has spread to more than 30 Provinces and Municipalities in various countries and regions as the center of Wuhan \cite{ref-1}. This outbreak not only poses a huge threat to people's health, but also has a huge economic and political damage, i.e., changing our daily life.

In order to control the epidemic, Wuhan (source of the virus, also Capital of Hubei Province) first announced to close the city on $24^{th}$ Jan, 2020 (Eve of Chinese New Year), then other Provinces and cities have started the first-class response to this major public health emergencies in succession to control population flow \cite{ref-3}. At the same time, medical teams carrying a large amount of medical supplies from other Provinces of China have arrived in Wuhan to support the local hospitals. However, the National epidemic situation, especially the situation in Hubei Province, is still disconcerting. The public is quite concerned about the development trend of the epidemic and expects the emergence of ``inflection point" \cite{ref-13}.

The outbreak of the COVID-19 has given rise to alarm for public health emergencies. So far, many medical problems about COVID-19  and its spread have yet to be studied, and questions and lessons are worthy investigating. Moreover, at the end of Spring Festival holiday, many enterprises will be back to work and the possibility of reemergence cannot be ignored and should be taken seriously.


Therefore, it is necessary to further study the epidemic process and the trend of the development of COVID-19, and provide a strong basis for effective control of disease epidemics. Focusing on the problem of work restoration, this paper establishes a mathematical model to evaluate the potential risk of returning to work during the epidemic period, and provide a theoretical basis for guiding enterprises  under safe and controllable condition.

In this paper, we first analyze the characteristics of the epidemic transmission from the visualization of the collected data, then establish the dynamic model of the infectious disease. Through the evaluation of the epidemic prevention and control measures, suggestions are provided considering the epidemic progression prediction, as a reference for epidemic prevention \& decision-making for public behaviours.

\section{Overview of the epidemic transmission in China}

\begin{figure}[!htp]
  \centering
  \includegraphics[width=0.9\linewidth]{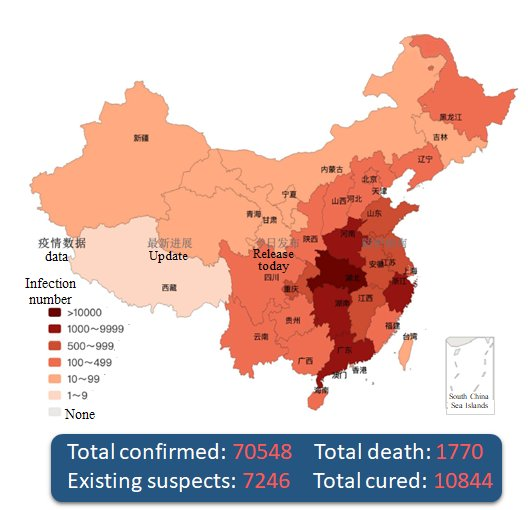}
  \caption{The heat map of COVID-19 in China (till 24:00 $17^{th}$ Feb 2020)}\label{fig-data-map}
\end{figure}

We use the daily confirmed data \cite{ref-0} published by the National Health Commission of the People's Republic of China (NHC-PRC) to reproduce the spread of NCP through the heat map, shown in Fig. \ref{fig-data-map}. It is easy to find that the spread of the epidemic is mainly centered in Wuhan and spread to the surrounding areas. Through the population mobility, the epidemic will spread to the central cities, i.e., Beijing, Shanghai, Guangzhou and other places, becoming the secondary transmission centers.

The source of the NCP is wild animal, reported by the Officials, and it can be transmitted via human-to-human, droplets or direct/indirect contact \cite{ref-6} \cite{ref-12}. There is no specific cure for the disease till now. On the other hand, the traditional Chinese medicine has made certain progress in clinical treatment to alleviate the symptom, and some experts pointed out that the plasma of cured patients has positive therapeutic effect. As a consequence, COVID-19 is included in category-B infectious disease by NHC-PRC, which is a contagious disease subject to quarantine, but China takes prevention and control measures corresponding to category-A infectious disease.

When the sudden outbreak of COVID-19 occurred at the New year eve of China, the Wuhan government failed to take effective measures in time to stop the spread of the NCP. Since Wuhan is the most important city in central China and transportation hub in China, it results in the rapid spread of the disease from Wuhan to the whole country, directly causing the epidemic control difficulty.

From $23^{rd}$ Jan 2020, Wuhan city was closed and stop all traffic with the outside, announced by the Government. A series of actions have been taken in order to reduce the risk of infection and cut the disease epidemic. Traffic control has been implemented all over China, and temperature checkpoints have been set up at highway junctions to check the vehicles. Furthermore, local governments implement community closed  management to restrict people travelling from the beginning of Feb 2020 in succession, especially in Hubei area and shuts down the entertainment and shopping malls. People are required to stay at home and not return to work and wait for further notice.

In the mean time, medical supplies and stuff are deployed and arranged from all over the country to support Wuhan and Hubei Province. Besides, the first two temporary hospitals, Leishenshan and Huoshenshan are constructed and completed in only 7 days to admit the patients. Till the midnight on $17^{th}$ Feb 2020, there were cumulatively reported 72,436 confirmed cases, 58,016 confirmed cases in mainland China, 6,242 suspected cases, cumulatively 1,868 death cases, 12,552 cured cases and discharged from hospital \cite{ref-0}. Total 92 confirmed cases were reported in Hong Kong, Macao and Taiwan, and 900 confirmed cases outside China, most of which are in countries near China. In particular, a total of 542 confirmed cases were reported on the Diamond Princess cruise which was docked at the Japanese dock.

\begin{figure}[!htp]
  \centering
  \includegraphics[width=0.9\linewidth]{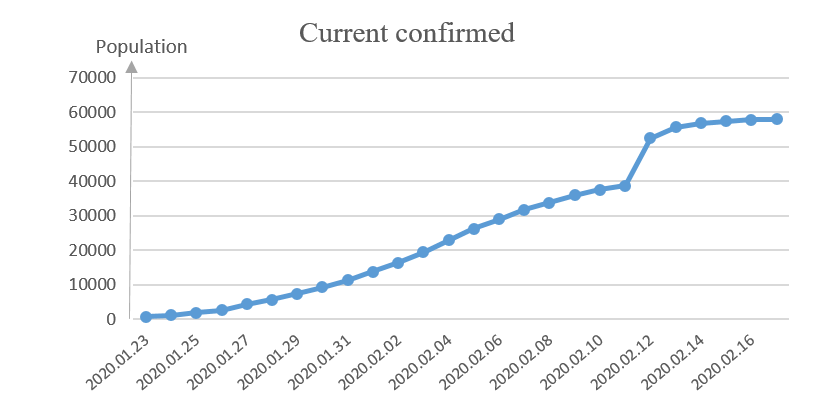}
  \caption{The current confirmed cases in China}\label{fig-x-1}
\end{figure}

\begin{figure}[!htp]
  \centering
  \includegraphics[width=0.9\linewidth]{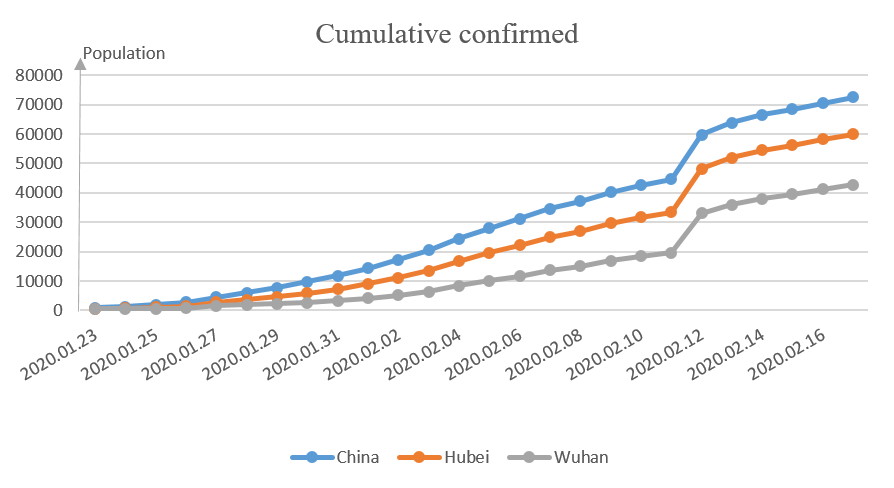}
  \caption{Comparison of cumulative confirmed cases in China, Hubei and Wuhan}\label{fig-x-2}
\end{figure}

\begin{figure}[!htp]
  \centering
  \includegraphics[width=0.9\linewidth]{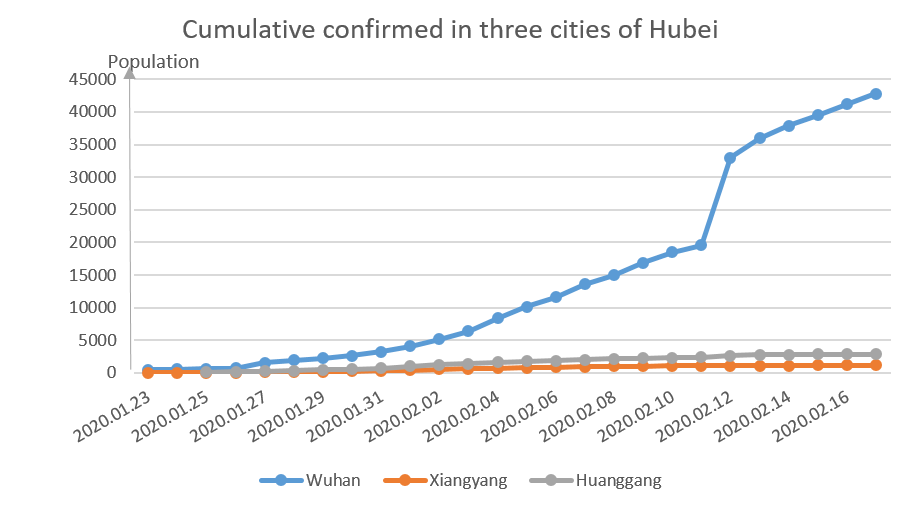}
  \caption{Comparison of cumulative confirmed cases among three cities in Hubei: Wuhan, Xiangyang and Huanggang}\label{fig-x-3}
\end{figure}

The current confirmed cases on daily in China is demonstrated in Fig. \ref{fig-x-1}, the cumulatively confirmed in China, Hubei Province and Wuhan city is demonstrated in  Fig. \ref{fig-x-2}, while the comparison of the three main areas in Hubei is demonstrated in Fig. \ref{fig-x-3}.



The data demonstrate that most of the confirmed cases are in Hubei Province, of which Wuhan is the most severely affected city. According to the reports, almost all confirmed cases outside Hubei province come from Hubei or have a history of contact with someone from Hubei. For instance, one of the authors who has infected NCP due to contact people from Wuhan on $22^{th}$ Jan 2020 at a gathering has just discharged from the hospital on $16^{th}$ Feb 2020.

Huanggang city and Xiangyang city are the most severe areas in Hubei Province, expect  Wuhan. It can be seen from Fig. \ref{fig-x-3} that the number of the confirmed cases in Wuhan is much larger than those in these two cities, which can also prove the effectiveness of closure of Wuhan from stopping the virus spread from $23^{rd}$ Jan, 2020.

\begin{figure}[!htp]
  \centering
  \includegraphics[width=0.9\linewidth]{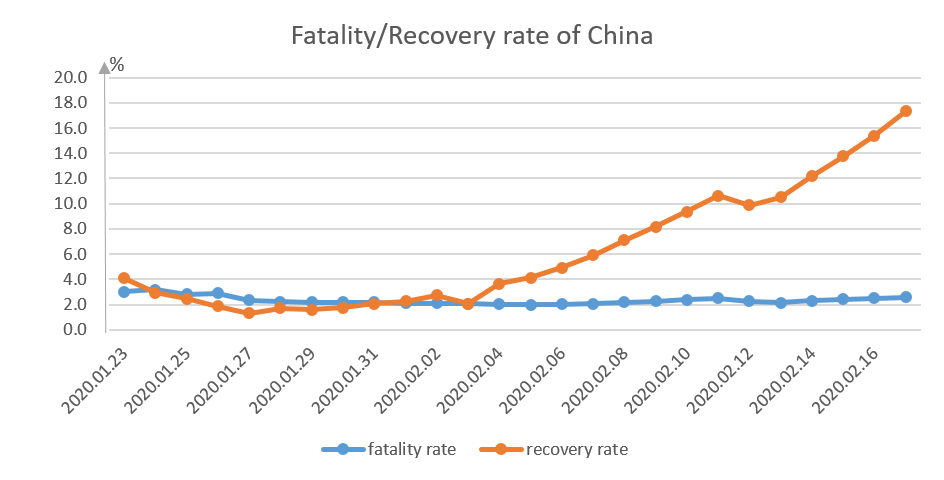}
  \caption{Comparison between the death rate and the cured rate}\label{fig-x-4}
\end{figure}
\begin{figure}[!htp]
  \centering
  \includegraphics[width=0.9\linewidth]{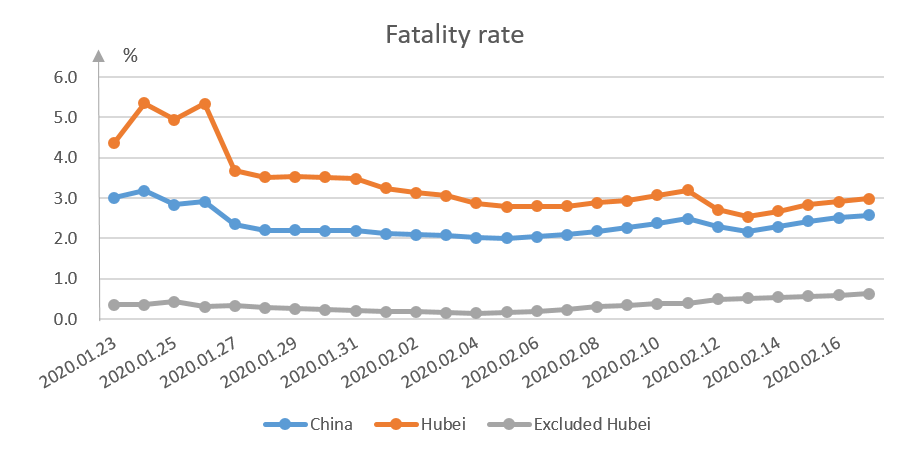}
  \caption{The tendency of the death rate in Hubei and other areas}\label{fig-x-5}
\end{figure}
\begin{figure}[!htp]
  \centering
  \includegraphics[width=0.9\linewidth]{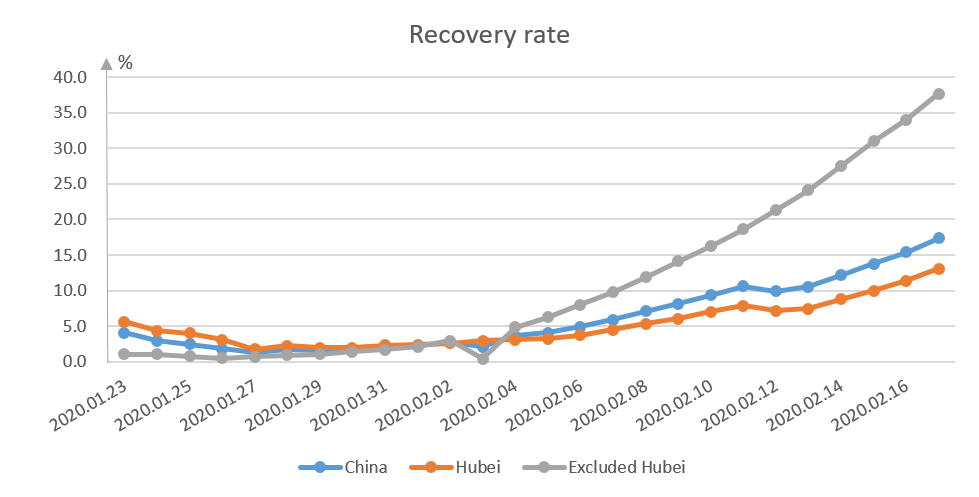}
  \caption{The tendency of the cured rate in China, Hubei and non-Hubei}\label{fig-x-6}
\end{figure}

\begin{figure}[!htp]
  \centering
  \includegraphics[width=0.9\linewidth]{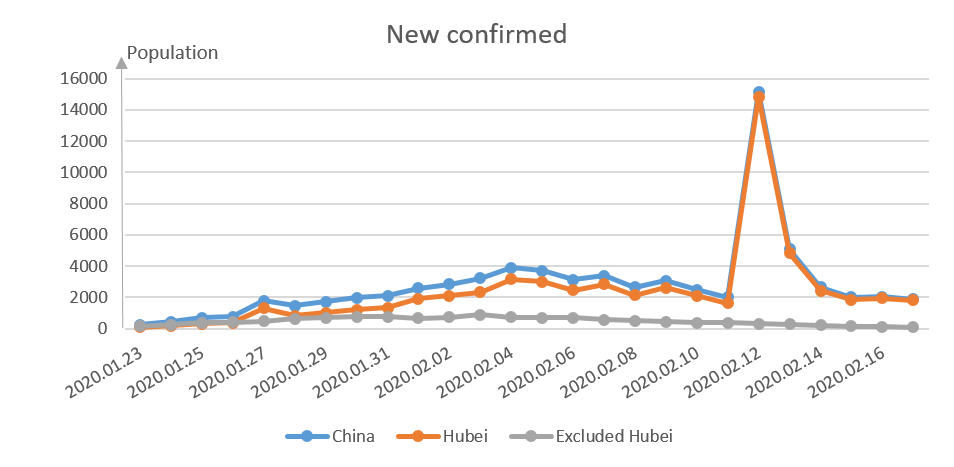}
  \caption{The new confirmed cases in China, Hubei and non-Hubei}\label{fig-x-7}
\end{figure}
\begin{figure}[!htp]
  \centering
  \includegraphics[width=0.9\linewidth]{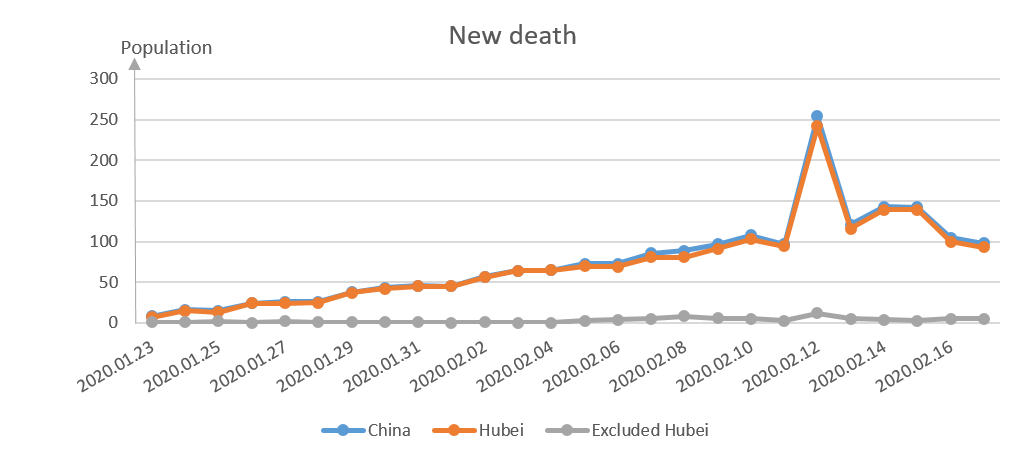}
  \caption{The new death cases in China, Hubei and non-Hubei}\label{fig-x-8}
\end{figure}

It can be seen from Fig. \ref{fig-x-4} that the cured rate has increased rapidly since the $3^{rd}$ Feb 2020, while the death rate is rather stable.
The tendency of the death rate and cured rate in Hubei and other areas are shown in Fig. \ref{fig-x-5} and Fig. \ref{fig-x-6}, repectively. As the origin of COVID-19, the death of Hubei Province is much higher than those of other Provinces, and the newly confirmed cases and newly dead cases are both much higher than the sum of other provinces, shown in Fig. \ref{fig-x-7} and Fig. \ref{fig-x-8}.

It demonstrates that the density of the infected people in Hubei Province is still quite large, and the current medical facilities cannot meet the demand. In contrast, the situation of the areas outside Hubei Province is rather easier to control due to effective community isolation and individual monitoring. Currently, different App programmes in each district or even community on the smart phone are developed to report personal tracks when on the move, i.e., `Shen i', `Health code'.
\begin{figure}[!htp]
  \centering
  \includegraphics[width=0.9\linewidth]{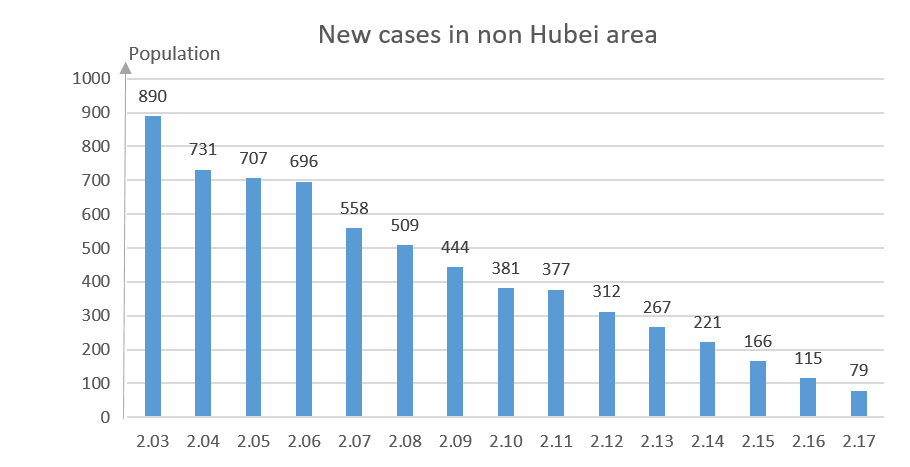}
  \caption{The tendency of new cases in non-Hubei area}\label{fig-x-10}
\end{figure}

As demonstrated in Fig. \ref{fig-x-7}, there is a sudden rise for the new confirmed cases in Hubei area at the $12^{th}$ Feb 2020, which is caused by the confirmation of the previous suspected population since the virus symptom can not be diagnosed at the first place \cite{ref-5} \cite{ref-10}. However, the inflected population with confirmation are roughly stable after this point, due to appropriate traffic and medical support from other Provinces of China. The new cases from non-Hubei area are decreasing gradually, shown in Fig. \ref{fig-x-10}. Therefore, it can be concluded that the current measures taken by the Chinese Government can effectively prevent the spread of virus.

\section{The YJ-SEIR model based propagation of COVID-19}

In the classical SEIR (Susceptible-Exposed-Infective-Recovered) model, the total population can be divided into four categories, i.e., susceptible population (S-type), referring to those who have no disease but lack of immune ability and are easy to be infected after contact with the infected ones; exposed (E-type), referring to the people who have been exposed to the infection but no obvious symptom in incubation period; infective (I-type), referring to the infected people who can transmit virus to the members of S-type and turning them to the  E-type or I-type members; recovered (R-type), referring the people  who are isolated or have immunity from the disease and they can be changed to S-type again if the immune period is limited \cite{ref-n-1}.

\begin{figure}[thbp]
\footnotesize
  \centering
  \includegraphics[width=0.8\linewidth]{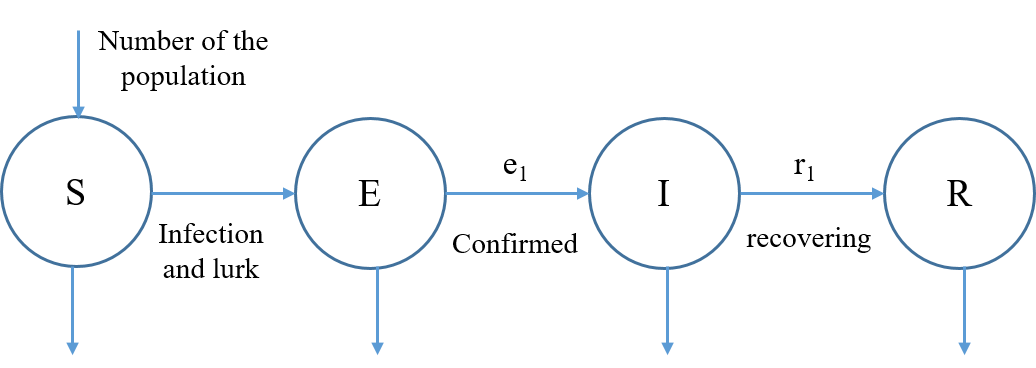}
  \caption{The SEIR state transferring model}\label{fig1-seir1}
\end{figure}

\begin{figure}[thbp]
  \centering
  \includegraphics[width=0.8\linewidth]{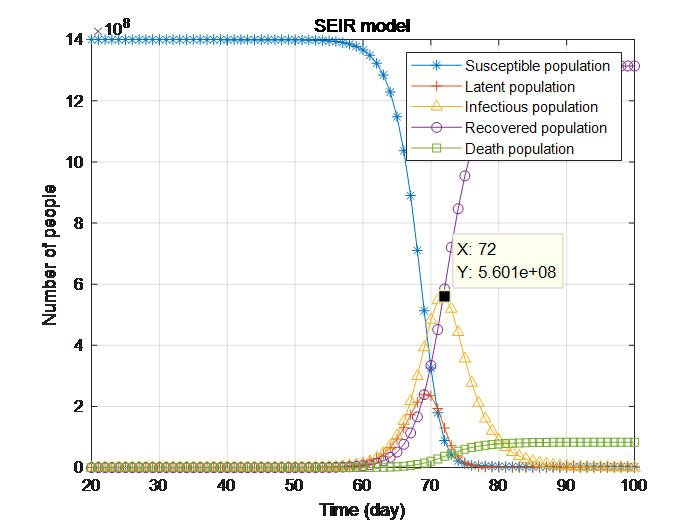}
  \caption{The prediction result based on SEIR model}\label{fig1-pre}
\end{figure}

Assuming these four types of people can be transferred among these categories based on certain probability, the state transition diagram and the prediction result are illustrated in Fig. \ref{fig1-seir1} and Fig. \ref{fig1-pre} as follows.

\begin{figure}
  \centering
  \includegraphics[width=0.9\linewidth]{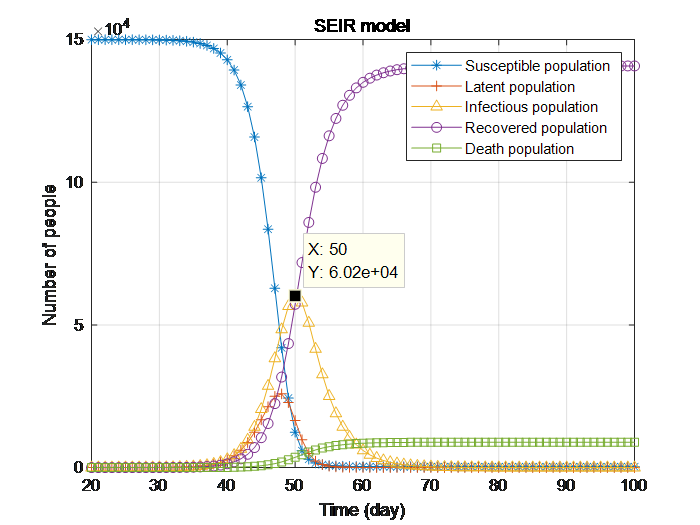}
  \caption{The SEIR model based prediction after isolation measures taken in Wuhan }\label{fig2}
\end{figure}

It can be seen from Fig. \ref{fig1-pre}, the peak of the infected people will appear around the $73^{rd}$ day with more than 56 million population without any isolation measures, which will be a huge disaster for  human beings.  In the case that the Chinese government adopts closing Wuhan and people are isolated at home such measures, however, only a small number of people who have to travel in the city, which can greatly reduce the base number of vulnerable people. We predict that there are 150,000 people travelling in the city  each day  and the estimation result is depicted in Fig. \ref{fig2}.

As shown in Fig. \ref{fig2}, the maximum number of the infected people reaches about 60,200 when the Chinese government take isolation measures against Wuhan. So the taken measures can effectively reduce the number of the infection. Besides, it takes 50 days to reach the inflection point (see Fig. \ref{fig2}), which is contradictory to the current virus development situation. In order to predict the COVID-19 progression more accurately, this paper proposes a YJ-SEIR model, which has been improved based on the typical SEIR model from three aspects:
\begin{enumerate}
  \item The backlog impact of the indirect infection and suspected cases on the development of the epidemic situation;
  \item The measures of medical staff from all over the country to support Wuhan and the construction of temporary rescue hospitals to improve the treatment capacity have positive impact on the suspected cases;
  \item The measures of traffic block and residential isolation.
\end{enumerate}

In the YJ-SEIR, 6 categories are divided for the population and two new categories are added, shown in Table \ref{tab1}.
\begin{table}[h]
    \caption{The new categories in the improved YJ-SEIR model}\label{tab1}
\scriptsize
  \centering
  \begin{tabular}{c|c}
    \hline
    Category & Meaning \\\hline
    \tabincell{c}{Y-type\\ (Suspected)} & \tabincell{l}{With symptom but not been confirmed in time} \\\hline
    \tabincell{c}{J-type \\(Indirectly infected)} &  \tabincell{l}{Indirectly infected people}  \\
    \hline
  \end{tabular}
\end{table}

\begin{figure}[thp]
  \centering
  \includegraphics[width=0.9\linewidth]{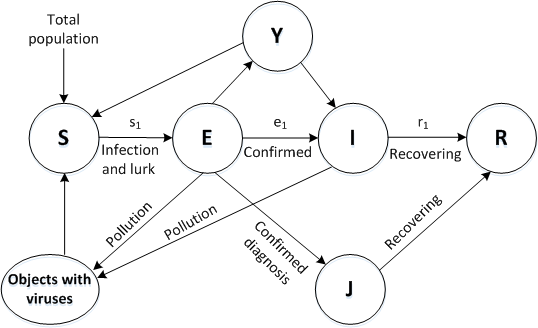}
  \caption{The state transferring model of the YJ-SEIR}\label{fig3-yg}
\end{figure}

\begin{figure}[thp]
  \centering
  \includegraphics[width=0.9\linewidth]{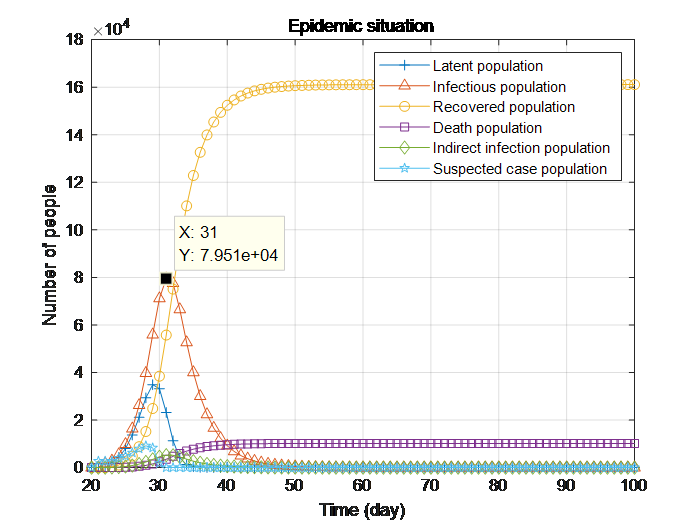}
  \caption{The prediction based on YJ-SEIR with isolation measures in Wuhan}\label{fig3-yg-pre}
\end{figure}

The state transferring diagram of the YJ-SEIR model is illustrated in Fig. \ref{fig3-yg}. Under this YJ-SEIR model, the virus prediction result of the Wuhan area is shown in Fig. \ref{fig3-yg-pre} without new rescue hospitals and any support from outside consideration, while the treatment capacity is 1000 cases and the recovery rate of the patients is set as 0.243.

It can be seen from Fig. \ref{fig3-yg-pre} that the number of infected people in Wuhan reaches the highest value of 79,510 in 31 days of closure, which is basically consistent with the current epidemic situation as reported. Chinese government has taken efficient measures to organize medical staff from all over the country to support Wuhan and construct temporary hospitals, which can increase hospital readiness and recovery rate.

\begin{figure}[h!]
    \centering
        \subfigure[with the hospital admission capacity and cure rate as (1000, 0.243)]{\includegraphics[width=0.9\linewidth]{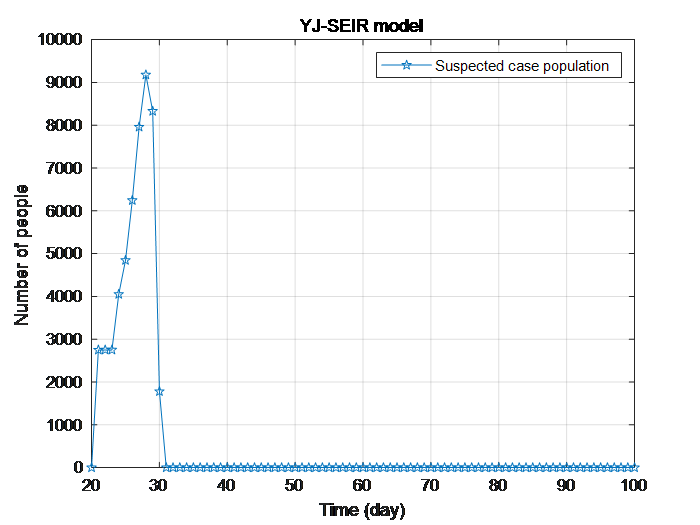}}\\
        \subfigure[with the hospital admission capacity and cure rate  as (5000, 0.243)]{\includegraphics[width=0.9\linewidth]{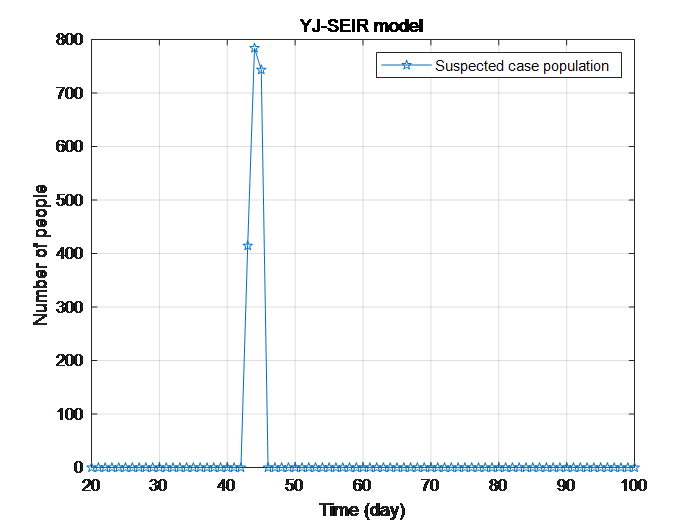}}\\
        \subfigure[with the hospital admission capacity and cure rate  as (15000, 0.3)]{\includegraphics[width=0.9\linewidth]{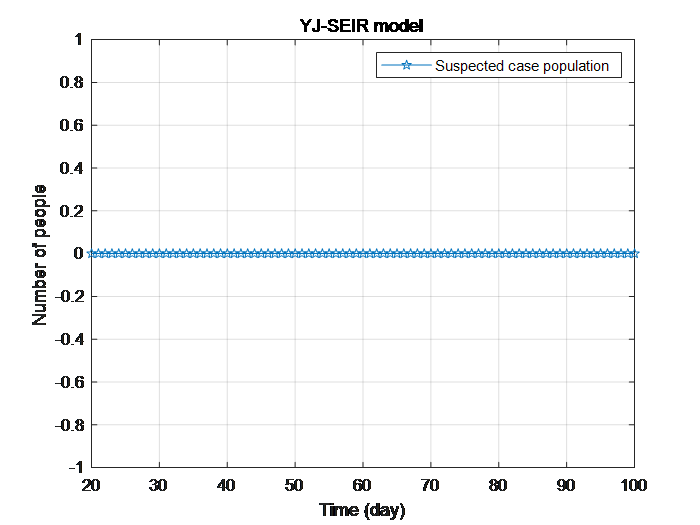}}\\
    \caption{The suspected population with YJ-SEIR model under taken measures}
    \label{fig-ca}
\end{figure}

Three groups of experiments are performed with the above measures taken by the Chinese government, and the experimental settings of the hospital admission capacity and recovery rate are set for each group as: (1000, 0.243), (5000, 0.243) and (15000, 0.3), respectively. The cumulative number of the suspected cases under these three situations is demonstrated in Fig. \ref{fig-ca}. It can be seen that the backlog of the suspected patients can be admitted timely and cured largely with enhanced hospital admission capacity.
%

\section{ Potential risk assessment of rework during epidemic period based on entropy-fuzzy factor}

The assessment of the potential risk of rework during COVID-19 epidemic period is a multi-level comprehensive evaluation problem, which involves various factors, especially  influenced by certain subjective factors \cite{ref-21}\cite{ref-22}\cite{ref-26}\cite{ref-27}. Therefore, it is necessary to distinguish the priority and determine the main factors. Based on the three basic components (i.e., source of infection, transmission route and susceptible population) of the epidemic of infectious diseases, the evaluation for the potential outbreak risk (EPOR) of the factory rework during epidemic mainly include such factors, listed in Table \ref{tab2}.


\begin{table*}[!htp]
  \caption{The EPOR of the work resumption during epidemic}\label{tab2}
\scriptsize
  \centering
  \begin{tabular}{p{1in}<{\centering}|c|p{1in}<{\centering}|c|c|c}
     \hline
    Self-protection & Company prevention & \tabincell{c}{ Protection \\supervision level}& \tabincell{c}{Production \\management level} & \tabincell{c}{Internal and external \\outbreak threat} & \tabincell{c}{Environmental \\impact}\\\hline
    \tabincell{l}{Wear disposable \\surgical masks;\\ Wear KN95 masks;\\ Separate eating;\\ Wash hands;\\ Wear KN95 masks \\and goggles;\\Wear Protective clothes;}
     & \tabincell{l}{Quarantine measure;\\
		Epidemic prevention; \\
		 Monitoring measure;\\
		Off-peak commuting; \\
		Perfect disinfection; \\
		Harmless treatment;}
       & \tabincell{l}{Perfecting epidemic \\prevention supervision \\system;\\
		Perfecting of quarantine \\supervision system; }
      &\tabincell{l}{ Density of workers \\at workplace;\\
		Worker communication \\frequency;\\
		Plant sector density;\\
		Management awareness of \\prevention and control; }
       & \tabincell{l}{Severity of the outside;\\
		External isolation strength;\\
		Epidemic trend;}
       & \tabincell{l}{Traffic condition;\\
		River abundance density;\\
		Bad weather;\\
		Mobility of \\contaminated objects;} \\
     \hline
   \end{tabular}
\end{table*}

As shown in Table \ref{tab2}, each factor is composed of several subfactors, which are either qualitative or quantitative. However, the risk of work restoration during epidemic can only be evaluated by the perfection degree. So the fuzzy comprehensive evaluation method is used here, i.e., entropy weight combined with fuzzy comprehensive analysis method \cite{ref-29}. The process of the EPOR model establishment are described as follows.

\textbf{Step 1) Determine the impact factor set of the EPOR for rework during epidemic.
}

The $1^{st}$-level indexes of EPOR for rework during epidemic are composed of 6 primary factors (see the first row in Table \ref{tab2}), and the $2^{nd}$-level indexes are composed of 25 risk factors (see the second row in Table \ref{tab2}),  written as $y=\{y_1, y_2, \cdots, y_{25}\}$, as shown in Table \ref{tab2}.

\textbf{Step 2) Determine the assessment set of EPOR for rework during epidemic.
}

The assessment set is first determined by the experts, where the assessment of each factor can be divided into $m$ levels, written as $V = \{V_1, V_2, \cdots, V_m\}$, denoting as $\{highest~risk,~ higher~risk,~ medium~risk,~ low~risk,~ lower$ $risk \}$. Here,  $m=5$ is selected.

\textbf{Step 3) Risk assessment based on entropy weight-fuzzy \cite{ref-28}.}

Let the fuzzy evaluation matrix for the EPOR for rework during epidemic be $R = \{r_{ij}\}{n\times m} , i = 1, 2, \cdots, n; j = 1, 2, \cdots, m$,
 where $r_{ij}$ is the membership degree of the $j^{th}$-level of the $i^{th}$ factor.
The ratio of the $j^{th}$ level with the $i^{th}$ index based on the entropy weight can be calculated as:
\begin{equation}\label{eq1}
  P_{ij}=(1+r_{ij})/\sum_{j=1}^{m}(1+r_{ij})
\end{equation}
Then the entropy of the $i^{th}$ index is calculated as:
\begin{equation}\label{eq2}
  e_{i}=-\sum_{j=1}^{m}{P_{ij}\bullet \ln P_{ij}/\ln m}
\end{equation}
So the entropy weight of the  $i^{th}$ index can be calculated as:
\begin{equation}\label{eq3}
  a_{i}=(1-e_i)/\sum_{i=1}^{n}{(1-e_i)}
\end{equation}
and the vector of the entropy weight of the EPOR of rework during epidemic can be written as $A=(a_1, a_2, \cdots, a_n)$. Therefore, the corresponding comprehensive evaluation vector $B$ is calculated as $B=A\times R$, denoted as $B = (b_1, b_2, \cdots, b_i, \cdots, b_m)$. There are many evaluation methods for the determination of $b_i$ considering the relationship among factors, such as principal subordinate relationship or balance all factors with weighted average model \cite{ref-29}.

\vspace{10pt}
\textbf{Case study}
\vspace{10pt}

According to the index listed in Table \ref{tab2}, a potential risk investigation for a factory to restore work randomly selected is conducted on the six primary indexes ($U_i, i=1,\cdots, 6$), and the statistical results are shown in Table \ref{tab3}.
\begin{table}[h]
 \caption{ Statistical table for major reprise risk evaluation}\label{tab3}
\footnotesize
  \centering
  \begin{tabular}{c|c|c|c|c|c}
    \hline
    \multirow{3}*{\tabincell{c}{The $1^{st}$-level \\index}} & \multicolumn{5}{c}{Risk assessment set} \\\cline{2-6}
      & \tabincell{c}{highest \\($V_1$)}  & \tabincell{c}{higher\\ ($V_2$)}  & \tabincell{c}{medium\\ ($V_3$)}  & \tabincell{c}{low\\ ($V_4$)} & \tabincell{c}{lower\\ ($V_5$)}  \\\hline
  $U_1$	&0.08&	0.17&	0.22&	0.29	&0.24\\\hline
$U_2$	&0.05&	0.16&	0.23&	0.31	&0.25\\\hline
$U_3$	&0.06&	0.18&	0.31&	0.21	&0.24\\\hline
$U_4$	&0.09&	0.16&	0.29&	0.27	&0.19\\\hline
$U_5$	&0.06&	0.18&	0.27&	0.32	&0.17\\\hline
$U_6$	&0.05&	0.14&	0.35&	0.27	&0.19\\
    \hline
  \end{tabular}
\end{table}

The  evaluation matrix is calculated in detail with the following steps.

1) The fuzzy evaluation matrix for obtaining the potential risk assessment of rework is:
\begin{equation}\label{eq4}
R = \left[ {\begin{array}{*{20}{c}}
{0.08}&{0.17}&{0.22}&{0.29}&{0.24}\\
{0.05}&{0.16}&{0.23}&{0.31}&{0.25}\\
{0.06}&{0.18}&{0.31}&{0.21}&{0.24}\\
{0.09}&{0.16}&{0.29}&{0.27}&{0.19}\\
{0.06}&{0.18}&{0.27}&{0.32}&{0.17}\\
{0.05}&{0.14}&{0.35}&{0.27}&{0.19}
\end{array}} \right]
\end{equation}

2) Calculate the rario of the index of the $j^{th}$ level under the $i^{th}$ index based on Eq. (\ref{eq1}):
\begin{equation}\label{eq5}
P = \left\{ {\begin{array}{*{20}{c}}
{0.1800}&{0.1950}&{0.2033}&{0.2150}&{0.2067}\\
{0.1750}&{0.1933}&{0.2050}&{0.2183}&{0.2083}\\
{0.1767}&{0.1967}&{0.2183}&{0.2017}&{0.2067}\\
{0.1877}&{0.1933}&{0.2150}&{0.2117}&{0.1983}\\
{0.1767}&{0.1967}&{0.2117}&{0.2200}&{0.1950}\\
{0.1750}&{0.1900}&{0.2250}&{0.2117}&{0.1983}
\end{array}} \right\}
\end{equation}

3) Calculate the entropy ($E$) of the $i^{th}$ index based on Eq.(\ref{eq2}):
\begin{equation}\label{eq6}
  E=\{0.9989~~0.9983~~ 0.9985 ~~   0.9988 ~ ~  0.9983 ~ ~  0.9977\}
\end{equation}

4) Calculate the entropy weight of the $i^{th}$ index according to formula (3):
\begin{equation}\label{eq7}
  A=\{0.1162   ~~ 0.1818  ~~  0.1548~ ~   0.1217  ~~  0.1829  ~ ~ 0.2427\}
\end{equation}

5) Calculate the fuzzy comprehensive evaluation using the weighted average model [2]:
\begin{equation}\label{eq7}
 B=A \times R=\{0.0900  ~ ~ 0.1800 ~~   0.2427   ~ ~0.2427~   ~ 0.1900\}
\end{equation}

Here, the risk assessment vector is set as $V=\{0.95,~0.75,~0.5,~0.35,~0.1\}$, so the comprehensive evaluation score of the potential risk assessment of a factory to restore work is calculated as, $D=B \times V^'$, thus the potential risk of resuming work at this factory is calculated to be 0.4423. In other words, the probability of the potential risk of this factory returning to work is about 44.23\%. Therefore, the head of the Enterprises can take necessary measures to restore normal production step by step.

\section{ Conclusion}
Based on the released public data of COVID-19 and the improved infectious disease dynamic model, the following comments are concluded:
\begin{enumerate}
  \item The Chinese government adopts five-level regional isolation measures in Wuhan, concentrates on the temporary isolation and rescue hospital construction, and organizes the medical staff of all Provinces from China to support Wuhan in improving their treatment capacity, so as to rapidly digest the backlog of suspected cases and reduce the number of viral infection.  It is of great significance that the implementation of these measures makes the virus infection rate in Wuhan well controlled.
  \item Considering the impact of indirect infection and suspected cases backlog on the epidemic situation, the prediction results can be more consistent with the progression of the epidemic.
  \item The human factors are not considered in the paper. Quite a few infected cases are caused by certain people wandering but ignoring close regulations during the epidemic or deliberately concealing contact history of Hubei.
\end{enumerate}

The epidemic is different from the general illness, which not only affects the people's health, but also affects the politics, economy, culture and education in China, as well as people's mental health and quality of life. In order to reduce the negative impact of the epidemic on whole development of China, it is urgent to effectively deal with the problem of the enterprise resumption and how to balance the epidemic prevention and economic development.

In this paper, the entropy and fuzzy comprehensive evaluation methods are combined to establish a potential risk assessment model for large-scale work resumption, so that such qualitative and fuzzy relationships related to the epidemic can be solved mathematically. As a result, the risk of work resumption under different protection conditions are evaluated by the model accurately and scientifically, which can provide theoretical guidance for the decision-making of the government and enterprises.

\section*{Acknowledgment}
This work was supported under National Natural Science Foundation of China (61973296).




\begin{thebibliography}{1}

\bibitem{ref-1} Chan J F W, Kok K H, Zhu Z, et al. Genomic characterization of the 2019 novel human-pathogenic coronavirus isolated from a patient with atypical pneumonia after visiting Wuhan. Emerging Microbes \& Infections, 2020, 9(1): 221-236.

\bibitem{ref-3} Tang B, Wang X, Li Q, et al. Estimation of the Transmission Risk of the 2019-nCoV and Its Implication for Public Health Interventions. Journal of Clinical Medicine, 2020, 9(2): 462.
\bibitem{ref-13} Zhao S, Lin Q, Ran J, et al. Preliminary estimation of the basic reproduction number of novel coronavirus (2019-nCoV) in China, from 2019 to 2020: A data-driven analysis in the early phase of the outbreak. International Journal of Infectious Diseases, 2020.
\bibitem{ref-0}http://www.nhc.gov.cn/
\bibitem{ref-6} Randhawa G S, Soltysiak M P M, El Roz H, et al. Machine learning-based analysis of genomes suggests associations between Wuhan 2019-nCoV and bat Betacoronaviruses. bioRxiv, 2020.
\bibitem{ref-12} Parry J. China coronavirus: cases surge as official admits human to human transmission. 2020.

\bibitem{ref-n-1}Kim S, Byun J H, Jung I H. Global stability of an SEIR epidemic model where empirical distribution of incubation period is approximated by Coxian distribution. Advances in Difference Equations, 2019, 2019 (1): 1 - 15.
\bibitem{ref-21} Miguel E, Chevalier V, Ayelet G, et al. Risk factors for MERS coronavirus infection in dromedary camels in Burkina Faso, Ethiopia, and Morocco, 2015. Eurosurveillance, 2017, 22(13).
\bibitem{ref-5} Yu F, Du L, Ojcius D M, et al. Measures for diagnosing and treating infections by a novel coronavirus responsible for a pneumonia outbreak originating in Wuhan, China. Microbes and Infection, 2020.
\bibitem{ref-10} Chen N, Zhou M, Dong X, et al. Epidemiological and clinical characteristics of 99 cases of 2019 novel coronavirus pneumonia in Wuhan, China: a descriptive study. The Lancet, 2020.


\bibitem{ref-22}Kim S W, Park J W, Jung H D, et al. Risk factors for transmission of Middle East respiratory syndrome coronavirus infection during the 2015 outbreak in South Korea. Clinical Infectious Diseases, 2017, 64(5): 551-557.
    \bibitem{ref-26} Sikkema R S, Farag E A B A, Himatt S, et al. Risk factors for primary Middle East respiratory syndrome coronavirus infection in camel workers in Qatar during 2013¨C2014: a case-control study. The Journal of infectious diseases, 2017, 215(11): 1702-1705.
\bibitem{ref-27} World Health Organization. Investigation of cases of human infection with Middle East respiratory syndrome coronavirus (MERS-CoV): interim guidance. World Health Organization, 2018.

\bibitem{ref-29} Liu K D , Pang Y J, Li W G, Jin Yan. Misleading Interpretation of Fuzzy Comprehensive Evaluation Weighted Average Model. Practice and understanding of mathematics, 2019, 49(08): 218-225.
\bibitem{ref-28} Feng A F, Cao P H. Assessment Model of Major Animal Epidemic Risks Based On Entropy Weight Fuzzy Comprehensive Evaluation. Journal of Animal Ecology, 2014, 035(008): 66-69.





%
%
%
%
%
%
%

%





\end{thebibliography}
%

\end{document}